\documentstyle[epsf]{article}%
\begin{document}

\title{Amino acid substitution matrices for protein conformation identification }
\author{Xin Liu${^1}$,Wei-Mou Zheng${^2}$\\
\small
${^1}${\it The Interdisciplinary Center of Theoretical Studies, Chinese Academy of Sciences, Beijing 100080, China}\\
\small${^2}${\it Institute of Theoretical Physics, China, Beijing
100080, China}}
\date{}
\maketitle

\begin{abstract}
Methods for alignment of protein sequences typically measure
similarity by using substitution matrix with scores for all
possible exchanges of one amino acid with another. Although widely
used, the matrices derived from homologous sequence segments, such
as Dayhoff's PAM matrices and Henikoff's BLOSUM matrices, are not
specific for protein conformation identification. Using a
different approach, we got many amino acid segment blocks. For
each of them, the protein secondary structure is identical. Based
on these blocks, we have derived new amino acid substitution
matrices. The application of these matrices led to marked
improvements in conformation segment search and homologues
detection in twilight zone.
\end{abstract}

\leftline{PACS number(s): 87.10.+e,02.50.-r}
\bigskip

\section{Introduction}
The similarity of amino acids is the basis of protein sequence
alignment, protein design, and protein structure prediction.The
mutation data matrices of Dayhoff\cite{pam} and the substitution
matrices of Henikoff\cite{blm} are standard choices of scores for
amino acid similarity evaluation. Although, widely used in protein
design and protein structure prediction, these matrices pay more
attention to homologous relationship than to conformation
similarity.

However, in studies of protein conformations, the task is to
detect whether residue sequences have similar conformations
neglecting their homologous relationships. Several
works\cite{bowie,pre67} showed that residue behavior is influenced
by protein conformations. Therefore, we wondered whether a better
approach might be to use alignments in which these relationships
are explicitly represented.

For this purpose, we derived many residue segment blocks from the
nonredundant PDB\_SELECT\cite{pslt} database. The amino acid
segments in each of the block have identical protein secondary
structure according to the database of secondary structure in
proteins(DSSP)\cite{dssp}. Consequently, based on the counts of
residue substitution in our database, we derived the amino acid
substitution matrices for protein conformation identification.

\section{Methods}
We created a nonredundant set of 1612 non-membrane proteins from
PDB\_SELECT with amino acid identity less than 25\% issued on 25
September of 2001. The secondary structure for these sequences
were taken from DSSP database. In DSSP algorithm, Kabsch and
Sander defined eight states of secondary structure according to
the hydrogen-bond pattern. As in most methods, we considered 3
states $\{h, e, c\}$ generated from the 8 by the coarse-graining
$H,G,I\to h$, $E\to e$ and $X,T,S,B\to c$.

\subsection{Constructing blocks databases}
For this work, we constructed blocks database from our dataset by
the following two rules:

(1) Each amino acid segment in a block has the same protein
secondary structure;

(2) Each amino acid segment in a block has at least one non-gapped
high scored alignment when it is compared with other segments in
the block.

\noindent{\bf Step 1:} A window of width $l$ is sliding along
every sequence of our dataset. A dataset $G$ of amino acid
segments $a_0a_1...a_{l-1}$ with the corresponding secondary
structure segments $s_0s_1...s_{l-1}$ is created. For an entry
$a^j_0a^j_1...a^j_{l-1}$ in dataset $G$, if there is another entry
$a^k_0a^k_1...a^k_{l-1}$ in $G$ which satisfies $s^j_0=s^k_0,
s^j_1=s^k_1, ...s^j_{l-1}=s^k_{l-1}$ and $S \geq S_0$
\begin{equation}
S=\sum^{i=l-1}_{i=0}score(a^j_i,a^k_i);
\end{equation}
then the entry is added to dataset $B'_{s_0s_1...s_{l-1}}$, where
$score(a^j_i,a^k_i)$ is the score for residues $a^j_i,a^k_i$
substitution in BLOSUM62\cite{blm} matrix. After filtering set
$\{B'\}$ to forbid sample recounts in the original dataset, we get
some high scored residue segments for different protein secondary
structure words.

\noindent{\bf Step 2:} To reduce the contributions to amino acid
pair frequencies from the less similar residue segments, segments
of each protein secondary structure word in set $\{B'\}$ are
clustered. This is done by specifying the threshold $S_0$,
identical to that used in step 1 , in which segments that have
alignment score $S$ better than $S_0$ are grouped together. For
example, in dataset $B'_{s_0s_1...s_{l-1}}$, if the score $S_{XY}
\geq S_0$ when segments $X$ and $Y$ are aligned, then $X$ and $Y$
are clustered. If segment $Z$ has a high scored alignment with
either $X$ or $Y$, it is also clustered with them.

After these two steps, we get a new blocks database fulfilling our
requirements.

\subsection{Deriving amino acid substitution matrices \\ from conformational blocks databases}
To reduce multiple contributions to amino acid pair  frequencies
from the most closely related members in a block, sequences are
clustered within blocks and each cluster is weighted as a single
sequence in counting pairs. This is done by specifying a
clustering percentage in which sequence segments that are
identical for at least that percentage of amino acids are grouped
together. In our work, we adjust this percentage to that of the
original BLOSUM matrices are derived.

We count all possible pairs of amino acid substitution in each
column of every block. All these counts are summed. The result of
this counting is a frequency table listing the number of times
each of the $20+19+..1=210 $ different amino acid pairs occurs
among the blocks. The table is used to calculate a matrix
representing the odds ratio between these observed frequencies and
those expected by chance.

We denote the total number of amino acid $i,j$ pairs($1\leq j\leq
i\leq 20 $) by $f_{ij}$. Then the observed probability of
occurrence for each $i,j$ pair is
\begin{equation}
q_{ij}=f_{ij}/\sum_{i=1}^{20}\sum_{j=1}^{i}f_{ij}
\end{equation}
The probability for the amino acid $i$ to occur is then
\begin{equation}
p_i=q_{ii}+\sum_{j\neq i}q_{ij}/2
\end{equation}
The expected probability of occurrence $e_{ij}$ for each $i,j$
pair is then $p_ip_j$ for $i=j$ and $p_ip_j+p_jp_i=2p_ip_j$ for
$i\neq j$. An odds ratio matrix is calculated where each entry is
$q_{ij}/e_{ij}$. A lod ratio is then calculated in bit units as
$s_{ij}=\log_2(q_{ij}/e_{ij})$. Lod ratios are multiplied by a
scaling factor of 2 and then rounded to the nearest integer value
to produce CBSM(conformational blocks substitution matrix)
matrices in half-bit units. For each substitution matrix, we
calculated the average mutual information per amino acid pair
$H$(also called relative entropy)\cite{rentroy}, and the expected
score $E$ in bit units as
\begin{equation}
H=\sum_{i=1}^{20}\sum_{j=1}^i q_{ij}\times s_{ij},   \qquad
E=\sum_{i=1}^{20}\sum_{j=1}^{20} p_i\times p_j\times s_{ij}.
\end{equation}

For more details on matrix driving, we refer reader to
Refs.\cite{blm}

\subsection{Protein secondary structure segment searches}
To evaluate the improvement of the performance of our amino acid
substitution matrices to that of the original BLOSUM matrix used,
we do a protein secondary structure segment search in the learning
set. All-against-all segment alignment is carried out in dataset
$G$. For two segments $X$ and $Y$, the alignment score $S_{XY}\geq
T_0$, if they have the same protein secondary structure then there
is a sample of Ture-Positive, else a False-Positive sample.

Given a threshold $T_{0B}$, using BLOSUM matrix, we get the counts
of Ture-Positive and False-Positive samples in dataset $G$. As our
substitution matrix is used, by varying the threshold $T_0$, we
adjust the counts of False-Positive samples to that of the
original BLOSUM matrix used. Consequently, we can evaluate the
improvement by the counts of Ture-Positive samples.

\subsection{Homologous pairs detection}
To evaluate the validity of the above scheme, we examine whether
the sequence alignments with CBSM60 perform better than those with
BLOSUM62 for homologous sequences in the twilight zone. For this
propose, we use the 176 sequences test set extracted by Elber et,
al\cite{s176}. Each homologous pair in the test set have a
sequence identity less than 25\%, but very similar structure.

We perform all-against-all sequence alignment on the test set with
Blast2.2.6\cite{blast1,blast2}. The gap insertion and elongation
parameters used for alignment are set to 11/1. Detective ability
is illustrated by the number of successfully identified homologous
pairs as a function of errors per query. The error per query is
defined as the total number of non-homologous protein sequences
detected with expectation value equal to or less than the
threshold divided by the total number of aligned sequence pairs.
By varying the expectation value cutoff of Blast, we get the
results shown in figure 1.

\section{Result}
By setting $S_0=27$, we created $l=10$ width blocks from our
dtatset. The resulting amino acid substitution matrices are shown
in table 1.

It is interesting to make a comparison between  the CBSM60
matrices obtained here with the commonly used BLOSUM62 matrices.
There are many remarkable differences between the two score
schemes. Lots of amino acid pairs have more negative score in
CBSM60 than those in BLOSUM matrices. For example, the scores for
residue pairs CN, CD, ID, WD, QC, KC, FC, WK, and WP are more than
two bits lesser than those in BLOSUM62 matrix. This means
dissimilar residues are strongly forbidden in amino acid
substitution. On the other hand, the scores for some pairs of
similar residues are slightly improved, such as SA, SN, LI, VI,
MI, ML, VL, FM, YF, and ST.

Relative entropy is 0 when the target(or observed) distribution of
pair frequencies is same as the background(or expected)
distribution and increases as these two distribution become more
distinguishable. Based on relative entropy, the BLOSUM90 is
comparable to CBSM60 with relative entropy of $\approx 1.2$ bit.
Some differences are seen when BLOSUM90 is subtracted from CBSM60
for every matrix entries. Compared to CBSM60, self substitution is
more preferable in BLOSUM90. For some amino acids, especially PW,
WD, and QC, CBSM60 is less tolerant to mismatches than BLOSUM90 is
. On the contrary, some substitutions, such as MA, FM, SN, VL, and
VY, are more tolerable in CBSM60.

The results of protein secondary structure segment searches are
shown in table 3. We find that there is a remarkable improvement
compared with those using BLOSUM matrix. The Ture-Positive sample
count for each $T_{0B}$ increases nearly 15\%. For False-positive
cases, the proportion of samples with tiny structural differences
between two aligned segments increases nearly one percent. This
means the CBSM substitution matrices work very well.

In the results of homologous pairs detection shown in figure 1, we
find that, compared with BLOSUM62, CBSM60 performs better. The
detected homologues increase nearly 1/3. Further more, for cases
where signal of sequence similarity is larger, we do the
homologues detection too. Compared with BLOSUM62, for SCOP40
database\cite{scop1,scop2} where only PDB sequences that have 40\%
homology or less are included, there is a slightly
increase($\approx $2\%) with CBSM60 for homologues detection.

Since the residues in each column of a block correspond to an
uniform secondary structure, we can get the residue pair counts
and calculate the amino acid substitution matrix for different
protein conformation states. The results are shown in table 4-6.
When the three matrices are compared with each other, we find many
differences. For example, comparing helix with sheet , the
similarities of CA, SR, MQ, PH, and TP change drastically. There
is a positive score for Cysteine and Alanine substitution in
helix. While in sheet conformation, the score is negative.

\section{Discussions}
We have found that substitution matrices based on amino acid pairs
in conformational blocks of aligned protein segments perform
better in protein secondary structure segment identification and
homologues detection than those based on Henikoff's BLOSUM score
scheme. Because the CBSM matrices can also be used in three
dimensional structure identification(This part will be published
elsewhere. ), the importance of such improved performance can be
profound for works in protein design and protein conformation
prediction.

Furthermore, CBSM is indeed a different scheme from BLOSUM. For
example, in the performance of homologues detection by CBSM60, we
found that 6 of 10 detected homologous pairs are different from
those detected using BLOSUM62 as the error per query equals 0.05.
When the error per query is 0.15, this portion is 19 of 31. This
means a new score scheme have been provided which can detect a
different scope of remote homologous relationship from Henikoff's
BLOSUM matrices.

There are fundamental differences between our approach and that of
Henikoff that could account for the superior performance of CBSM
matrices. In their case,based on Prosite database, blocks were
derived primarily from the most highly conserved regions of
proteins in residue sequence means neglecting the conformation
identity. Many of the differences between CBSM and BLOSUM matrices
may arise from multi-conformation regions of conserved sequence.

Our results show the strong dependence of residue behavior on
conformations. From table 4-6, we find that there are many residue
pairs displaying strong dependence of similarity on conformations.
We expect that specially derived scores for multiple conformations
should work better in researches of protein structure. This will
be discussed elsewhere.

\begin{quotation}
{This work is part of the project 10347145 supported by National
Natural Science Foundation of China.}
\end{quotation}

\newpage
Table 1. CBSM60 substitution matrix(Lower) and difference
matrix(Upper) obtained by subtracting the BLOSUM62 matrix position
by position. \small
\begin{verbatim}
    A   R   N   D   C   Q   E   G   H   I   L   K   M   F   P   S   T   W   Y   V
    1   0  -1  -1  -1   0  -1  -1   0  -1   0   0   1   0  -1   1   0  -3   0   0  A
        1   0  -1  -2   0   0  -1   0  -1  -1   0   0   0  -1   0   0  -3  -1  -1  R
A   5       0   0  -4   0  -1   0   0  -2  -1   0  -3  -1  -2   1   0  -2  -2  -2  N
R  -1   6       0  -5   0   0  -1  -1  -4  -3  -1  -2  -2  -2   0  -1  -6  -3  -3  D
N  -3   0   6       0  -5  -3  -1  -2  -1  -1  -4  -1  -4  -2  -2  -1  -3  -2  -2  C
D  -3  -3   1   6       0   0  -1   0  -1   0   0  -1  -3  -3   0   0  -3  -1  -2  Q
C  -1  -5  -7  -8   9      -1  -2   0  -2  -2   0  -1  -3  -2   0  -1  -2  -2  -2  E
Q  -1   1   0   0  -8   5       1  -2  -3  -1  -1  -1  -2   0   0  -1  -2  -2  -1  G
E  -2   0  -1   2  -7   2   4       0  -2   0   0  -1  -1  -2   0  -1  -3  -1  -1  H
G  -1  -3   0  -2  -4  -3  -4   7       0   1  -1   1   0  -3  -1   0  -2  -1   1  I
H  -2   0   1  -2  -5   0   0  -4   8       0  -1   1   0  -3  -1   0   0  -1   1  L
I  -2  -4  -5  -7  -2  -4  -5  -7  -5   4       0  -1  -1  -2   0   0  -4  -1  -1  K
L  -1  -3  -4  -7  -2  -2  -5  -5  -3   3   4       0   1  -2   0  -1  -1   0   0  M
K  -1   2   0  -2  -7   1   1  -3  -1  -4  -3   5       1   0  -1  -1  -1   1   0  F
M   0  -1  -5  -5  -2  -1  -3  -4  -3   2   3  -2   5       0  -1  -1  -7  -3  -2  P
F  -2  -3  -4  -5  -6  -6  -6  -5  -2   0   0  -4   1   7       0   1   0  -1  -2  S
P  -2  -3  -4  -3  -5  -4  -3  -2  -4  -6  -6  -3  -4  -4   7       1  -2  -1   0  T
S   2  -1   2   0  -3   0   0   0  -1  -3  -3   0  -1  -3  -2   4      -1  -1  -3  W
T   0  -1   0  -2  -2  -1  -2  -3  -3  -1  -1  -1  -2  -3  -2   2   6       0   0  Y
W  -6  -6  -6 -10  -5  -5  -5  -4  -5  -5  -2  -7  -2   0 -11  -3  -4  10       0  V
Y  -2  -3  -4  -6  -4  -2  -4  -5   1  -2  -2  -3  -1   4  -6  -3  -3   1   7
V   0  -4  -5  -6  -3  -4  -4  -4  -4   4   2  -3   1  -1  -4  -4   0  -6  -1   4
    A   R   N   D   C   Q   E   G   H   I   L   K   M   F   P   S   T   W   Y   V
\end{verbatim}
\newpage
\normalsize Table 2. The difference matrix between CBSM60 and
BLOSUM90 obtained by subtracting the BLOSUM90 matrix position by
position. \small
\begin{verbatim}
A   0
R   1   0
N  -1   1  -1
D   0   0   0  -1
C   0   0  -3  -3   0
Q   0   0   0   1  -4  -2
E  -1   1   0   1  -1   0  -2
G  -1   0   1   0   0   0  -1   1
H   0   0   1   0   0  -1   1  -1   0
I   0   0  -1  -2   0   0  -1  -2  -1  -1
L   1   0   0  -2   0   1  -1   0   1   2  -1
K   0   0   0  -1  -3   0   1  -1   0   0   0  -1
M   2   1  -2  -1   0  -1   0   0   0   1   1   0  -2
F   1   1   0   0  -3  -2  -1   0   0   1   0   0   2   0
P  -1   0  -1   0  -1  -2  -1   1  -1  -2  -2  -1  -1   0  -1
S   1   0   2   1  -1   1   1   1   1   0   0   1   1   0   0  -1
T   0   1   0   0   0   0  -1   0  -1   0   1   0  -1   0   0   1   0
W  -2  -2  -1  -4  -1  -2   0   0  -2  -1   1  -2   0   0  -6   1   0  -1
Y   1   0  -1  -2   0   1   0   0   0   0   0   0   1   1  -2   0  -1  -1  -1
V   1  -1  -1  -1  -1  -1  -1   1   0   1   2   0   1   1  -1  -2   1  -3   2  -1
    A   R   N   D   C   Q   E   G   H   I   L   K   M   F   P   S   T   W   Y   V
\end{verbatim}
\newpage

Table 3. Results of protein secondary structure segment searches.\\
\begin{tabular}{|l|l|c|ccccccccccc|}\hline\hline
& & \multicolumn{11}{c|}{ Protein secondary structure differences
}&\\ \cline{3-14}

& &0&1&2&3&4&5&6&7&8&9&10&\\ \cline{4-13}

$T_{0B}$&&($T_P$) &\multicolumn{10}{c}{ Percentage of
$F_p$}&$F_p$\\\hline
  &CBSM60   &101745&\bf{7.8}&9.2 &10.5&11.0&11.3&11.3&10.7&9.7&7.7&10.7&1368517\\
23&BLOSUM62 &86601 &6.9     &8.7 &10.5&11.4&12.0&12.0&11.3&9.9&7.6& 9.7&1389714\\\hline
  &CBSM60   &46999 &\bf{8.1}&9.5 &10.8&11.1&11.4&11.3&10.6&9.5&7.5&10.2&597618\\
25&BLOSUM62 &39522 &7.2     &9.0 &10.8&11.6&12.1&12.0&11.1&9.7&7.3& 9.2&588929\\\hline
  &CBSM60   &20612 &\bf{8.5}&9.9 &11.0&11.3&11.5&11.3&10.5&9.3&7.2& 9.6&242292\\
27&BLOSUM62 &17841 &7.6     &9.3 &11.0&11.8&12.2&12.0&10.9&9.5&7.0& 8.6&243169\\\hline
  &CBSM60   &9272  &\bf{8.9}&10.3&11.2&11.4&11.5&11.3&10.3&9.0&6.9& 9.1&98926\\
29&BLOSUM62 &8175  &8.1     &9.7 &11.3&11.8&12.4&12.0&10.6&9.1&6.8& 8.1&98219\\\hline
  &CBSM60   &4356  &\bf{9.7}&10.7&11.4&11.6&11.6&11.2& 9.9&8.7&6.7& 8.5&39101\\
31&BLOSUM62 &3942  &8.9     &10.2&11.5&11.9&12.5&11.9&10.1&9.0&6.7& 7.4&38688\\\hline
 \hline
\end{tabular}\\

\begin{figure}
\centerline{\epsfxsize=12cm \epsfbox{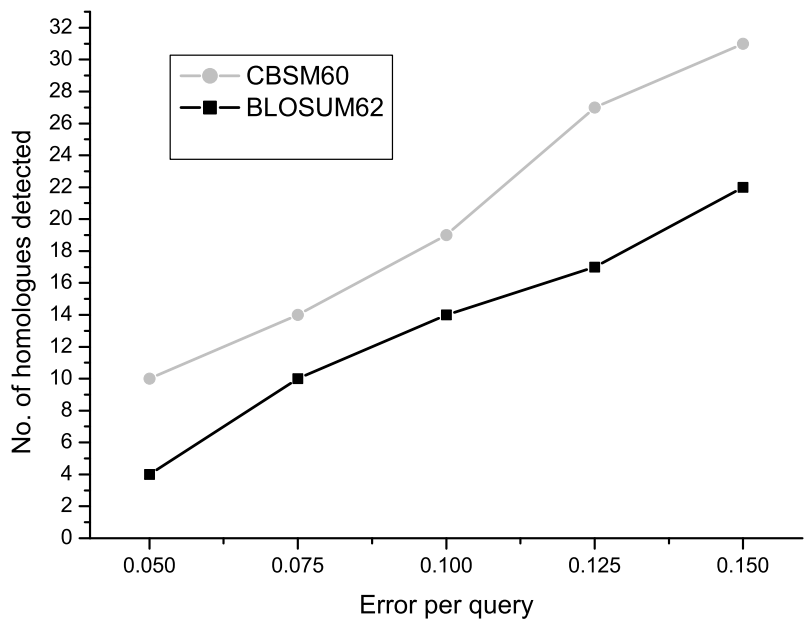}} \caption{ The number
of successfully identified homologous pairs in the 176 sequences
test set as a function of errors per query.} \label{fig1}
\end{figure}
\newpage      \normalsize
Table 4. Amino acid substitution matrix CBSM60c for coil
state(Lower) and difference matrix(Upper) obtained by subtracting
the CBSM60h matrix position by position. The bold entries are
pairs which have different positive/negative signs in the two
compared matrices.
\\
\begin{tabular}{cccccccccccccccccccccc}
 &  A&  R&  N&  D&  C&  Q&  E&  G&  H&  I&  L&  K&  M&  F&  P&  S&  T&  W&  Y&  V&   \\
 & -1&  1& -1& -1& \bf{-3}&  0&  2&  0&  1&  1&  2&  2&  1&  0&  1&  0&  0&  2&  2&  1& A \\
 &   &  1&  0&  0&  1&  1&  0& -1& -2&  2&  2&  1&  1&  4& -3&  1&  0&  1&  4&  3& R \\
A&  4&   & -2&  0&  1&  0& -1& \bf{-2}&  0&  3&  1& -1&  2&  2& -2&  0& -1&  7&  0&  3& N \\
R&  0&  6&   & -1&  2& -1&  0& -2&  0&  1&  1&  0&  1&  5&  1& -1& -1&  4&  0& -1& D \\
N& -3&  0&  5&   & -3& -4&  2&  2& -4&  1&  0& -1& -1& -5& -6& -4& -2&  7& -3& -3& C \\
D& -3& -3&  1&  5&   &  1&  1& -2&  0&  2&  3&  1& \bf{3}&  5&  1&  0&  1& -1&  3&  3& Q \\
C& -2& -4& -7& -8&  8&   &  1& -1&  0&  4&  2&  1&  3&  4&  2&  0&  1&  1&  1&  3& E \\
Q& -1&  2&  0& -1&-11&  5&   & -4& -3& -1& -1&  0&  2&  1&  1& \bf{-2}& -3&  1& -1&  1& G \\
E&  0&  0& -1&  2& -6&  3&  5&   &  1& -1&  1&  1&  4&  0&  2& -1&  0&  4&  1&  4& H \\
G&  0& -3& -1& -3& -5& -3& -3&  5&   &  2&  1&  0&  1&  0&  2&  2&  2&  4&  1&  1& I \\
H& -1& -2&  1& -2& -7&  0&  0& -5&  9&   &  1&  1&  1&  1&  1&  1&  0& -2&  0&  1& L \\
I& -1& -2& -3& -6&  0& -3& -2& -6& -6&  5&   &  0&  2&  1&  1&  0&  1&  1&  3&  1& K \\
L&  0& -1& -3& -6& -1&  0& -4& -5& -3&  4&  5&   &  1&  1& -3&  1&  3& -3&  0&  1& M \\
K&  0&  3& -1& -2& -8&  2&  1& -3& -1& -4& -2&  5&   &  1&  3&  2&  0& \bf{3}&  1&  1& F \\
M&  0&  0& -3& -4& -2&  1& -1& -2& -1&  3&  4&  0&  6&   & -6&  0& -1& -3&  4&  3& P \\
F& -3& -1& -3& -3& -8& -3& -3& -4& -2&  0&  1& -3&  1&  8&   &  0&  0&  3&  2&  2& S \\
P& -1& -4& -5& -3& -7& -3& -2& -4& -4& -5& -4& -2& -4& -3&  5&   & -1&  2&  1&  1& T \\
S&  2&  0&  1& -1& -5&  0&  0& -1& -2& -1& -2&  0&  0& -2& -3&  4&   &  1&  0&  4& W \\
T&  0& -1&  0& -2& -3&  0& -1& -4& -3&  0& -1& -1&  0& -4& -3&  2&  5&   &  1& -1& Y \\
W& -5& -5& -3& -7& -3& -7& -4& -3& -4& -2& -4& -6& -5&  2&-11& -2& -2& 11&   &  1& V \\
Y& -1& -1& -4& -5& -5&  0& -3& -5&  2& -1& -2& -1& -1&  5& -5& -2& -2&  1&  8&   &   \\
V&  1& -2& -3& -6& -4& -2& -2& -3& -2&  5&  3& -2&  2&  0& -2& -2&  0& -2& -2&  5&   \\
 &  A&  R&  N&  D&  C&  Q&  E&  G&  H&  I&  L&  K&  M&  F&  P&  S&  T&  W&  Y&  V&   \\
\end{tabular}
\newpage
Table 5. Amino acid substitution matrix CBSM60e for sheet
state(Lower) and difference matrix(Upper) obtained by subtracting
the CBSM60c matrix position by position. The bold entries are
pairs which have different positive/negative signs in the two
compared matrices.
\\
\begin{tabular}{cccccccccccccccccccccc}
 &  A&  R&  N&  D&  C&  Q&  E&  G&  H&  I&  L&  K&  M&  F&  P&  S&  T&  W&  Y&  V&   \\
 &  1& -1&  2&  0& -1&  1& -1&  1&  0& -1& -1& -1&  0&  1& -4&  0&  0& -1& -3& -1& A \\
 &   &  0&  1&  1&  0&  1&  2&  0&  2& -3& -3&  0& -1& -2&  3&  1&  0& -2& -1& -1& R \\
A&  5&   &  3& -1&  2&  2& \bf{2}& \bf{2}& -1& -6& -1& \bf{2}& -1& -4&  2&  2&  0& -3&  1& -3& N \\
R& -1&  6&   &  4&  2& \bf{2}&  0&  1&  1& -1& -1&  1& -1& -5& -1& \bf{2}& -3&  1& -3&  0& D \\
N& -1&  1&  8&   &  2&  8& -1& -2&  2& -4& -2&  3& -2&  0&  4& -1&  0& -1& -3&  1& C \\
D& -3& -2&  0&  9&   &  1&  0&  2&  1&  0& -4&  0&  1& -3&  0&  0& -1&  5& -3& -2& Q \\
C& -3& -4& -5& -6& 10&   &  1&  2&  0& -2& -3&  1& -5& -3&  2&  0&  1& -1& -2& -2& E \\
Q&  0&  3&  2&  1& -3&  6&   &  3& -3&  0&  0&  0& -3& -1&  0& \bf{2}&  0& -4&  0&  0& G \\
E& -1&  2&  1&  2& -7&  3&  6&   & -1&  3&  1&  0& -3& -1& \bf{5}&  2&  2&  1& -1& -2& H \\
G&  1& -3&  1& -2& -7& -1& -1&  8&   & -3& -2&  0& -2& -2&  2& -2& -2& -6& -3& -3& I \\
H& -1&  0&  0& -1& -5&  1&  0& -8&  8&   & -1& -2& -1& -1& -3& -1& -1&  2&  0& -2& L \\
I& -2& -5& -9& -7& -4& -3& -4& -6& -3&  2&   &  1& -2& -1& -3&  1&  0& -2& -4& -2& K \\
L& -1& -4& -4& -7& -3& -4& -7& -5& -2&  2&  4&   & -1& -1&  2& -1& -2&  0& -2& -1& M \\
K& -1&  3&  1& -1& -5&  2&  2& -3& -1& -4& -4&  6&   & -2& -2&  0&  2& \bf{-3}& -2& -2& F \\
M&  0& -1& -4& -5& -4&  2& -6& -5& -4&  1&  3& -2&  5&   &  6&  1& \bf{4}&  7& -1& -6& P \\
F& -2& -3& -7& -8& -8& -6& -6& -5& -3& -2&  0& -4&  0&  6&   &  1&  0& -5&  0& -4& S \\
P& -5& -1& -3& -4& -3& -3&  0& -4&  1& -3& -7& -5& -2& -5& 11&   &  0& -3& -2& -1& T \\
S&  2&  1&  3&  1& -6&  0&  0&  1&  0& -3& -3&  1& -1& -2& -2&  5&   & -2& -1& -9& W \\
T&  0& -1&  0& -5& -3& -1&  0& -4& -1& -2& -2& -1& -2& -2&  1&  2&  5&   & -2& -1& Y \\
W& -6& -7& -6& -6& -4& -2& -5& -7& -3& -8& -2& -8& -5& -1& -4& -7& -5&  9&   & -2& V \\
Y& -4& -2& -3& -8& -8& -3& -5& -5&  1& -4& -2& -5& -3&  3& -6& -2& -4&  0&  6&   &   \\
V&  0& -3& -6& -6& -3& -4& -4& -3& -4&  2&  1& -4&  1& -2& -8& -6& -1&-11& -3&  3&   \\
 &  A&  R&  N&  D&  C&  Q&  E&  G&  H&  I&  L&  K&  M&  F&  P&  S&  T&  W&  Y&  V&   \\
\end{tabular}
\newpage
Table 6. Amino acid substitution matrix CBSM60h for helix
state(Lower) and difference matrix(Upper) obtained by subtracting
the CBSM60e matrix position by position. The bold entries are
pairs which have different positive/negative signs in the two
compared matrices.
\\
\begin{tabular}{cccccccccccccccccccccc}
 &  A&  R&  N&  D&  C&  Q&  E&  G&  H&  I&  L&  K&  M&  F&  P&  S&  T&  W&  Y&  V&       \\
 &  0&  0& -1&  1& \bf{4}& -1& -1& -1& -1&  0& -1& -1& -1& -1&  3&  0&  0& -1&  1&  0& A \\
 &   & -1& -1& -1& -1& -2& -2&  1&  0&  1&  1& -1&  0& -2&  0& \bf{-2}&  0&  1& -3& -2& R\\
A&  5&   & -1&  1& -3& -2& -1&  0&  1&  3&  0& -1& -1&  2&  0& -2&  1& -4& -1&  0& N     \\
R& -1&  5&   & -3& -4& -1&  0&  1& -1&  0&  0& -1&  0&  0&  0& -1&  4& -5&  3&  1& D     \\
N& -2&  0&  7&   &  1& -4& -1&  0&  2&  3&  2& -2&  3&  5&  2&  5&  2& -6&  6&  2& C     \\
D& -2& -3&  1&  6&   & -2& -1&  0& -1& -2&  1& -1& \bf{-4}& -2& -1&  0&  0& -4&  0& -1& Q\\
C&  1& -5& -8&-10& 11&   & -2& -1&  0& -2&  1& -2&  2& -1& -4&  0& -2&  0&  1& -1& E     \\
Q& -1&  1&  0&  0& -7&  4&   &  1&  6&  1&  1&  0&  1&  0& -1&  0&  3&  3&  1& -1& G     \\
E& -2&  0&  0&  2& -8&  2&  4&   &  0& -2& -2& -1& -1&  1& \bf{-7}& -1& -2& -5&  0& -2& H\\
G&  0& -2&  1& -1& -7& -1& -2&  9&   &  1&  1&  0&  1&  2& -4&  0&  0&  2&  2&  2& I     \\
H& -2&  0&  1& -2& -3&  0&  0& -2&  8&   &  0&  1&  0&  0&  2&  0&  1&  0&  0&  1& L     \\
I& -2& -4& -6& -7& -1& -5& -6& -5& -5&  3&   & -1&  0&  0&  2& -1& -1&  1&  1&  1& K     \\
L& -2& -3& -4& -7& -1& -3& -6& -4& -4&  3&  4&   &  0&  0&  1&  0& -1&  3&  2&  0& M     \\
K& -2&  2&  0& -2& -7&  1&  0& -3& -2& -4& -3&  5&   &  1& -1& -2& -2&  0&  1&  1& F     \\
M& -1& -1& -5& -5& -1& -2& -4& -4& -5&  2&  3& -2&  5&   &  0& -1& \bf{-3}& -4& -3&  3& P\\
F& -3& -5& -5& -8& -3& -8& -7& -5& -2&  0&  0& -4&  0&  7&   & -1&  0&  2& -2&  2& S     \\
P& -2& -1& -3& -4& -1& -4& -4& -5& -6& -7& -5& -3& -1& -6& 11&   &  1&  1&  1&  0& T     \\
S&  2& -1&  1&  0& -1&  0&  0&  1& -1& -3& -3&  0& -1& -4& -3&  4&   &  1&  1&  5& W     \\
T&  0& -1&  1& -1& -1& -1& -2& -1& -3& -2& -1& -2& -3& -4& -2&  2&  6&   &  1&  2& Y     \\
W& -7& -6&-10&-11&-10& -6& -5& -4& -8& -6& -2& -7& -2& -1& -8& -5& -4& 10&   &  1& V     \\
Y& -3& -5& -4& -5& -2& -3& -4& -4&  1& -2& -2& -4& -1&  4& -9& -4& -3&  1&  7&   &       \\
V&  0& -5& -6& -5& -1& -5& -5& -4& -6&  4&  2& -3&  1& -1& -5& -4& -1& -6& -1&  4&       \\
 &  A&  R&  N&  D&  C&  Q&  E&  G&  H&  I&  L&  K&  M&  F&  P&  S&  T&  W&  Y&  V&       \\
\end{tabular}

\end{document}